\documentclass[aps,prl,twocolumn]{revtex4-1}
\usepackage{amssymb}
\usepackage{epsfig}
\usepackage{graphicx}
\usepackage{amsmath}
\usepackage{amsfonts}
\usepackage{mathrsfs}

\newcommand{\id}{{\sf 1 \hspace{-0.3ex} \rule{0.1ex}{1.52ex} \rule[-.02ex]{0.3ex}{0.1ex} }}

\begin{document}

\title{Computation of 2-D Spectra Assisted by Compressed Sampling}

\author{J. Almeida$^{1,2}$, J. Prior$^{3}$ and M.B. Plenio$^{1,2}$}
\address{$^{1}$ Institute for Theoretical Physics, Albert-Einstein-Allee 11, University Ulm, D-89069 Ulm, Germany}
\address{$^{2}$ Institute for Integrated Quantum Science and Technology, Albert-Einstein-Allee 11, University Ulm, D-89069 Ulm, Germany}
\address{$^{3}$ Departamento de F\'{i}sica Aplicada, Universidad Polit´ecnica de Cartagena, 
Cartagena 30202, Spain}

\begin{abstract}
The computation of scientific data can be very time consuming even if they are ultimately
determined by a small number of parameters. The principle of compressed sampling suggests
that for typical data we can achieve a considerable decrease in the computation
time by avoiding the need to sample the full data set. We demonstrate the usefulness
of this approach at the hand of 2-D spectra in the context of ultra-fast non-linear
spectroscopy of biological systems where numerical calculations are highly challenging
due to the considerable computational effort involved in obtaining individual data points.
\end{abstract}

\maketitle

{\em Introduction --} Signals of interest that are obtained in
experiments or by numerical computation of natural phenomena are not
white noise but contain hidden structure and redundancy. The fact that
it is possible to make use of these structures is the basis of
compressed sampling \cite{CandesW08,CandesRT06,CandesT06,Donoho06}
which demonstrates, surprisingly, that the accurate reconstruction of
such signals can be achieved with high probability from a number of
data points that would be deemed insufficient by the Nyquist - Shannon
criterion. This promise of considerable efficiency gains together with
the development of numerical algorithms that allow for its efficient
implementation have led to an explosion of applications of compressed
sampling in signal processing \cite{Review}.

Here we apply the principle of compressed sampling to the problem of
non-linear two-dimensional (2-D) spectroscopy and demonstrate that its
application can offer a considerable reduction in the computational
effort involved in the numerical determination of such spectra. This
makes accessible 2-D spectra for larger systems or for systems that
are governed by more complex dynamical equations, for example due to
the inclusion of non-Markovian features
\cite{Ishizaki09,PriorCHP10,ChinRHP10,KreisbeckK12}.  This
computational saving is not restricted to 2-D spectra. It serves
however to highlight the observation that numerical computation of
signals arising in nature can be made considerably more efficient
employing the paradigm of compressed sampling. This is true whenever
the signal is determined by a small number of parameters but the
computation of individual data points is very costly
\cite{AlanI,AlanII}. It is noteworthy that the same ideas may also be
applied to reduce the experimental effort in measuring 2-D
spectra. Here however the impact of experimental noise on the quality
of this reconstruction requires careful consideration \cite{CandesP10}
and will be studied elsewhere.

{\em Principles --} In non-linear 2-D spectroscopy a series of three
(or more) laser pulses is used to probe the dynamics of a system and
the resulting information is arranged as a matrix representing the two
dimensional Fourier transform of the signal which in turn describes
the system response in dependence of the spacing between laser pulses
(more details will be given in the next section). The point by point
computation of the entries of this matrix can be very time consuming,
especially when the relevant dynamical system is high-dimensional or
subject to a complex non-Markovian system-environment interaction. As
explained above, we expect however that this matrix contains
structure. In line then with the paradigm of compressed sampling it
will be possible to achieve a considerable reduction in the number of
elements of the signal that need to be obtained through (random)
sampling (by experimental observation or numerical computation). After
an outline of the numerical algorithm that is being used we provide a
demonstration of the achievable computational gain at the hand of a
specific bio-molecular complex of current interest.

Let $M$ be an $n\times n$ matrix that represents the 2-D spectrum that we wish to reconstruct
and let experimental observation or numerical computation determine a small subset of these elements
described by the set of indices  $\Omega$. Denoting by $tr|X|$ the trace norm of a matrix $X$
it can be proven that the solution of the minimization problem
\begin{equation}
    \min[tr|X|\, : X_{ij}=M_{ij}\; \mbox{for}\; (i,j)\in \Omega] \label{tracenorm}
\end{equation}
is {\em unique} and yields the matrix $M$ with a probability larger than $1-exp(-\beta)$ if the
number of sampled entries of $M$ is of order $nr(1+\beta)\ln n$, where $\beta>0$ is an arbitrary
constant and $r$ is the rank of the matrix $M$ (see \cite{CandesR08,Gross11} for proofs and a
rigorous mathematical statement). While reconstruction of the matrix $M$ cannot be guaranteed
with certainty, the probability for obtaining the correct reconstruction of the matrix $M$ can
be increased arbitrarily by repeating the sampling and reconstruction procedure $k$ times.
Accepting the majority result increases the probability that the so obtained result is correct
to above $1-e^{-k\beta}$. For typical values of $n$ in the range of $100 - 1000$ only $\beta=\ln n$
this reduces the probability of a wrong reconstruction to $10^{-8} - 10^{-12}$. Given that the
computation time for the full data set scales as $n^2$ this suggest that a considerable potential
reduction in computational effort is possible.

The solution of the minimization problem eq. \eqref{tracenorm} can be
obtained by standard solvers for semidefinite optimization problems
which do however tend to be limited to relatively small matrix sizes
with several hundred elements only. Here we solve the minimization
problem of eq. \eqref{tracenorm} by means of the so-called singular
value thresholding (SVT) algorithm \cite{CaiCS08,CramerPFSGBLPL10}
which permits very large matrices to be treated and to provably
approximate the solution of eq. \eqref{tracenorm} to arbitrary
precision. The SVT-algorithm solves iteratively the set of equations
\begin{align}
    &Y^{(k-1)}= U^{(k-1)}D^{(k-1)} V^{(k-1)}\label{SVT1}\\
    &X^{(k)} = U^{(k-1)}\max(D^{(k-1)}-\tau\id,0)V^{(k-1)}\label{thresh}\\
    &Y^{(k)} = Y^{(k-1)} + \delta {\cal P}_{\Omega}(M-X^{(k)})\label{SVT}
\end{align}
where ${\cal P}_{\Omega}(M)$ is a matrix whose entries satisfy $({\cal
  P}_{\Omega}(M))_{ij}=M_{ij}$ for $(i,j)\in \Omega$ and are zero
otherwise while $\tau$ and $\delta$ are constants whose choice will be
described below. The algorithm is initialized by a random matrix
$Y^{(0)}$ for the first iteration $k=1$. Then the singular value
decomposition of $Y^{(0)}$ is computed in eq. \eqref{SVT1}. This is
followed by the so-called soft thresholding step eq. \eqref{thresh}
where we subtract from all the singular values the parameter $\tau$
and set negative results to $0$. This step yields a matrix $X^{(k)}$
that has lower rank than $Y^{(k-1)}$. The final step of iteration $k$
then uses $X^{(k)}$ to construct a matrix $Y^{(k)}$ that satisfies the
constraints $X_{ij}=M_{ij}\; \mbox{for}\; (i,j) \in \Omega$ more
closely than the matrix $Y^{(k-1)}$ eq. \eqref{SVT}. These steps are
iterated until convergence has been achieved.

The choice $\delta<2$ ensure provable convergence of the iteration. For $\tau$ the choice $\tau=5n$
for $n\times n$-matrices leads to fast convergence and very close approximation to the solution of
eq. \eqref{tracenorm} (see \cite{CaiCS08} for details). Making use of the sparseness of the matrices
the SVT algorithm is capable of treating very large matrices
(reaching $30000\times 30000$ and above). Variants of this algorithm can
achieve a considerable additional increase in computational efficiency \cite{MaGC08,BeckerBC11}
by employing linear time algorithms for the implementation of the singular value decomposition
in eq. \eqref{SVT1}.

Whenever the calculation of individual data points is costly, it will be advantageous
to apply the compressed sampling paradigm and compute only a small subset of the matrix
elements and complete the remaining entries of the matrix with the SVT algorithm. In the
following we would like to demonstrate the power of this approach in the calculation
of 2-D spectra.

{\em Applications --} Non-linear spectroscopy
has proven to be useful in order to unveil the dynamics involved in
excitonic transfer of light harvesting complexes, due to the fact that
it is sensitive to excitonic quantum superpositions, i.e., excitonic
coherences
\cite{Mukamel95,Cho09,EngelCRAMCBF07,PanitchayangkoonHFCHWBE10} as
well as vibrational features of the protein environment
\cite{HayesPFCWFE10,CaycedoSolerCAHP12}.  Non-linear 2-D spectroscopy
can resolve the third-order polarization of the electronic system,
from the signal $S^{(3)}(t_1, t_2, t_3)$, arising from
photo-excitation of three consecutive pulses with wave vectors
$\textbf{k}_1$, $\textbf{k}_2$ and $\textbf{k}_3$, separated by time
intervals $t_1$ and $t_2$ and measured at a time $t_3$ after the
last pulse. It is customary to Fourier transform the time $t_1$ and
$t_3$ dimensions to yield $S(\omega_1, t_2, \omega_3)$ in order to
generate a 2-D spectra parametrized by the waiting time $t_2$. The
numerical computation as well as the experimental determination of
these 2-D spectra represent a challenging task because of the large
number of measurements that need to be taken and the considerable
computational resources required for the determination of
$S^{(3)}(t_1, t_2, t_3)$.

Here we will demonstrate by means of numerical examples, that the paradigm of
compressed sampling delivers a considerable gain in computational efficiency
as it is not necessary to determine the full signal $S^{(3)}(t_1, t_2, t_3)$
for a given choice of $t_2$. The computation for a randomly chosen subset of
times $t_1,t_3$ suffice to reconstruct the full signal.\newline

We consider the 2-D spectra of a pigment-protein complex, the
Fenna-Matthews-Olson (FMO) complex, which has received considerable
attention recently \cite{PanitchayangkoonHFCHWBE10}. In a spectroscopy
experiment the pigment-protein complex dynamics is probed by means of
applying different laser pulses and measuring the response to
them. Using Liouville space notation we can describe the evolution of
our system under the effect of the laser pulses by the master equation
\begin{equation}
    \frac{d\rho}{dt} = -\frac{i}{\hbar}\mathscr{L}\rho
- \frac{i}{\hbar}\mathscr{L}_{\rm int}(t)\rho.
\end{equation}
Here the Liouvillian $\mathscr{L}$ describes the internal excitation
dynamics including dephasing of the complex while the term
$\mathscr{L}_{\rm int}(t)$ stands for the interaction with the
laser. Typically in these complexes the dissipation time scales are
much longer than the dephasing induced by the environment and hence we
will consider only pure dephasing contributions to the internal
dynamics. That is, we will consider an unperturbed Liouvillian with
the form $\mathscr{L}\equiv \mathscr{L}_{\rm Ham} + \mathscr{L}_{\rm
  deph}$ . Within the Born-Markov approximation we can write each of
the terms in the equations above as:
\begin{align}
&\mathscr{L}_{\rm Ham}\equiv [H_s,\rho]\\
&\mathscr{L}_{\rm deph}\equiv i\sum_{i=1}^7 2\gamma_i(\sigma_i^z\rho\sigma_i^z - \rho)\\
&\mathscr{L}_{\rm int}(t)=[H_{\rm int}(t), \rho],
\label{eq:Ldef}
\end{align}
with $\sigma_x$, $\sigma_y$, $\sigma_z$ the standard Pauli matrices
and $H_{int}=-\vec{E}(t)\cdot \vec{V}$. For the homogeneous dephasing
rates we have chosen random values within a realistic interval for
each chromophore, i.e., $\{\gamma_i,\}_{i=1..7} = \{0.94, 2.26, 2.83,
1.70, 1.70, 1.88, 2.07\}\times 10^{-3} {\rm rad/fs}$ (with the
conversion $1{\rm rad}\cdot{\rm fs}^{-1}/1{\rm
  cm}^{-1}=2\pi\cdot2.99792458\times 10^{-5}$), which correspond to dephasing
times in the scale of few picoseconds. As regards the actual values of
the site energies and intersite coupling rates in the one-exciton
sector we have used the following Hamiltonian\cite{CarusoPRB12} (in
units of $\rm{rad}/\rm{fs}$):

\begin{widetext}

\begin{equation}
H_s=\left(
\begin{array}{rrrrrrr}
    0.0405 & -0.0224 &  0.0013 & -0.0014 &  0.0016 & -0.0034 & -0.0028 \\
   -0.0224 &  0.0575 &  0.0070 &  0.0018 &  0.0004 &  0.0031 &  0.0013 \\
    0.0013 &  0.0070 &       0 & -0.0131 & -0.0003 & -0.0022 &  0.0007 \\
   -0.0014 &  0.0018 & -0.0131 &  0.0377 & -0.0147 & -0.0040 & -0.0143 \\
    0.0016 &  0.0004 & -0.0003 & -0.0147 &  0.0801 &  0.0208 & -0.0012 \\
   -0.0034 &  0.0031 & -0.0022 & -0.0040 &  0.0208 &  0.0593 &  0.0081 \\
   -0.0028 &  0.0013 &  0.0007 & -0.0143 & -0.0012 &  0.0081 &  0.0499 \\
\end{array}
\right)
+2.297\cdot \mathbf{1}.
\label{eq:Hsite}
\end{equation}

\end{widetext}

Finally, the dipolar operator is defined from the individual dipole
momenta of each chromophore $\{\vec{\mu}_i\}_{i=1\dots 7}$ as
$\vec{V}=\sum_{i=1}^7 \vec{\mu}_i \sigma^x_i$. The directions of the
seven induced-transition dipoles $\vec{\mu}_i$, extracted from the
structure of the FMO complex \cite{TronrudPR09}, are

\begin{equation}
\frac{[\mu_i]}{\vert \vec{\mu}_{\rm BCh} \vert}=
\left(
\begin{array}{rrr}
-0.026 &  0.286 & -0.958\\
-0.752 &  0.601 & -0.271\\
-0.935 &  0.061 &  0.349\\
-0.001 &  0.393 & -0.919\\
-0.739 &  0.672 &  0.048\\
-0.859 &  0.371 & -0.353\\
-0.176 & -0.042 & -0.983\\
\end{array}
\right)
\label{eq:Dipoles}
\end{equation}

and their magnitude is taken to be the same and given in units of the
dipole strength $\vert \vec{\mu}_{\rm BCh} \vert$ of an isolated
bacteriochlorophyll.

The signal measured on a general spectroscopic experiment is
intimately related to the system electric response functions of
different orders $S^{(n)}$. In particular in non-linear 2-D
spectroscopy experiments the signal can be related to the
third-order response function $S^{(3)}(t_1, t_2, t_3)$. This function
can be written in a very compact form using tetradic notation in
Liouville space \cite{Mukamel95}
\begin{equation}
S^{(3)}(t_1, t_2, t_3)=\Big(\frac{i}{\hbar}\Big)^3
\langle\langle V\vert \mathscr{G}(t_3)\mathscr{V}
\mathscr{G}(t_2)\mathscr{V}
\mathscr{G}(t_1)\mathscr{V}
\vert \rho(-\infty)\rangle \rangle
\label{eq:S3def}
\end{equation}
where $\mathscr{G}(t)$ is the Liouville space Green Function in the
absence of the radiation field
\begin{equation}
    \mathscr{G}(t)\equiv \theta(t)\exp \Big( \frac{i}{\hbar} \mathscr{L}t\Big)
\end{equation}
and the action of the superoperator $\mathscr{V}$ upon an
ordinary operator $A$ is defined as
\begin{equation}
    \mathscr{V}A\equiv [V, A].
\end{equation}

\begin{figure*}[t]
\centering
$\begin{array}{cc}
\includegraphics[scale=0.6]{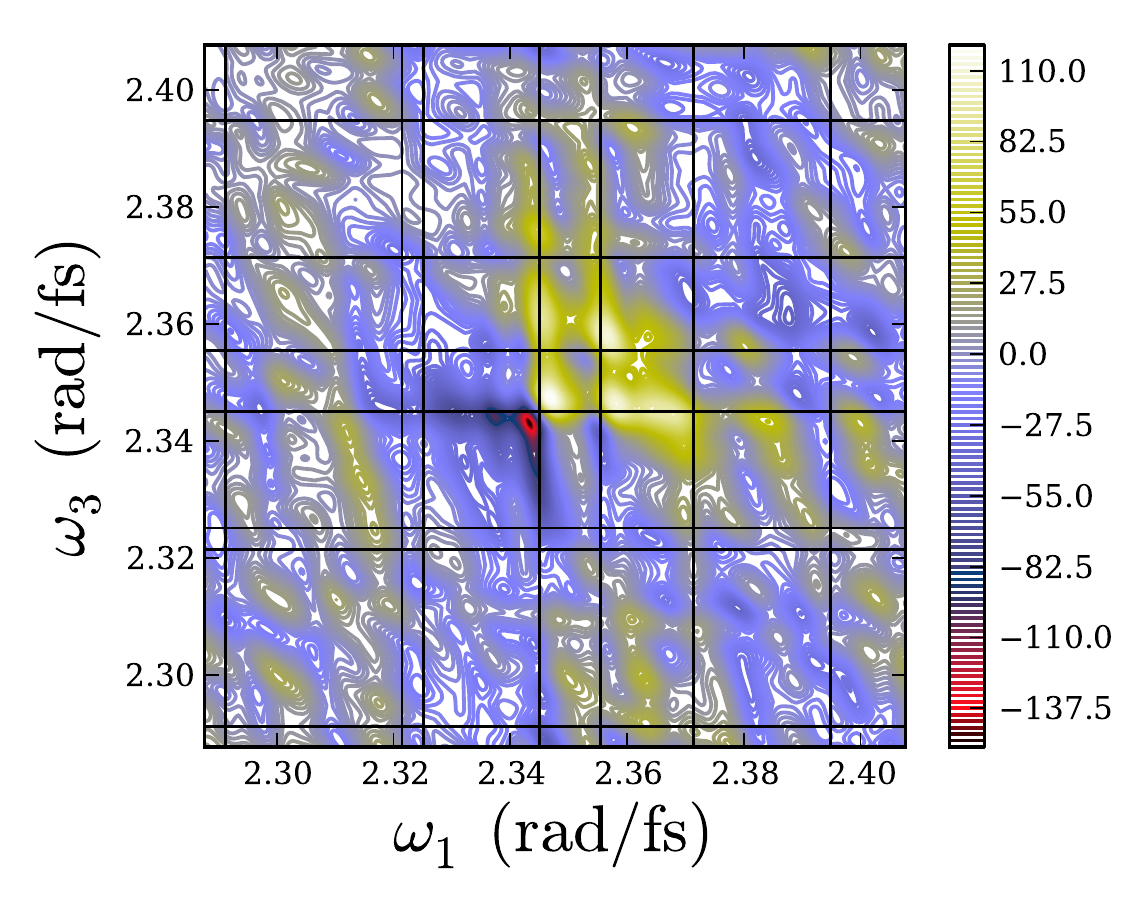} &  \includegraphics[scale=0.6]{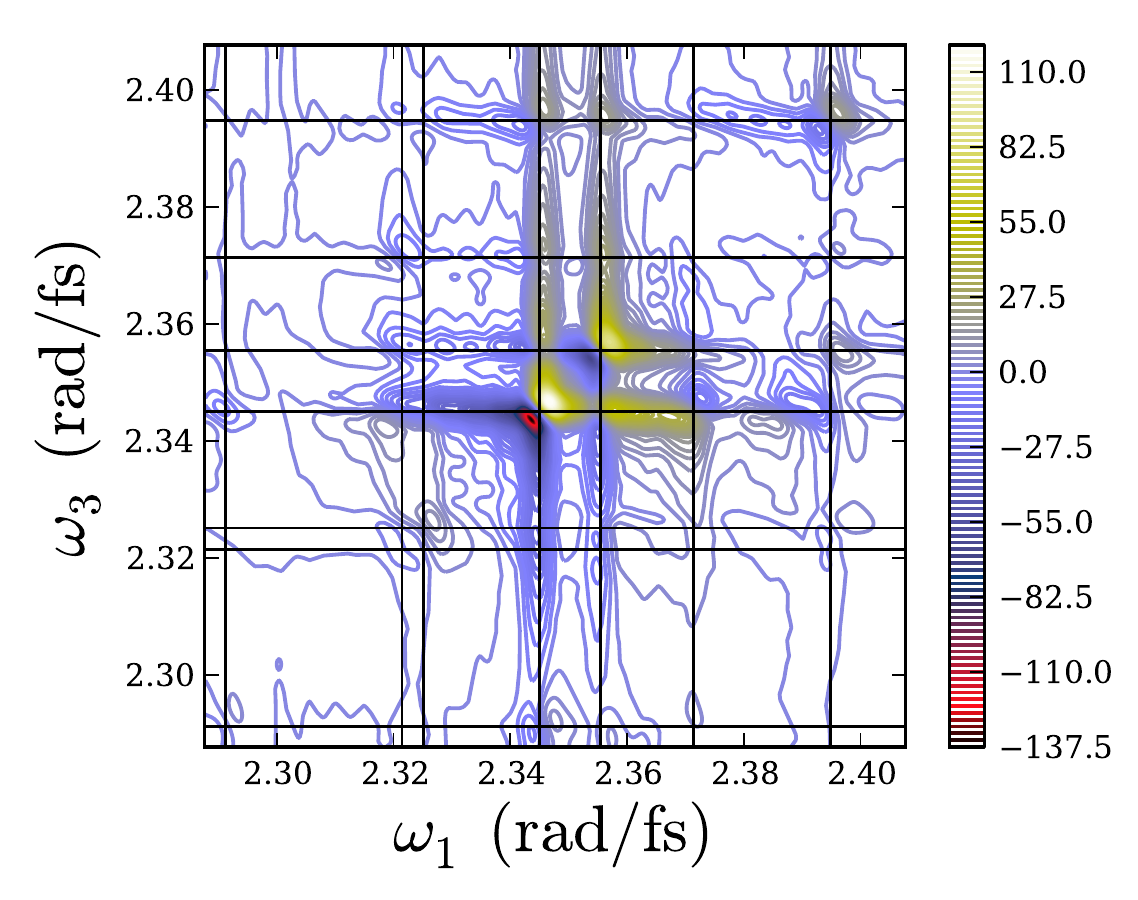} \\
\includegraphics[scale=0.6]{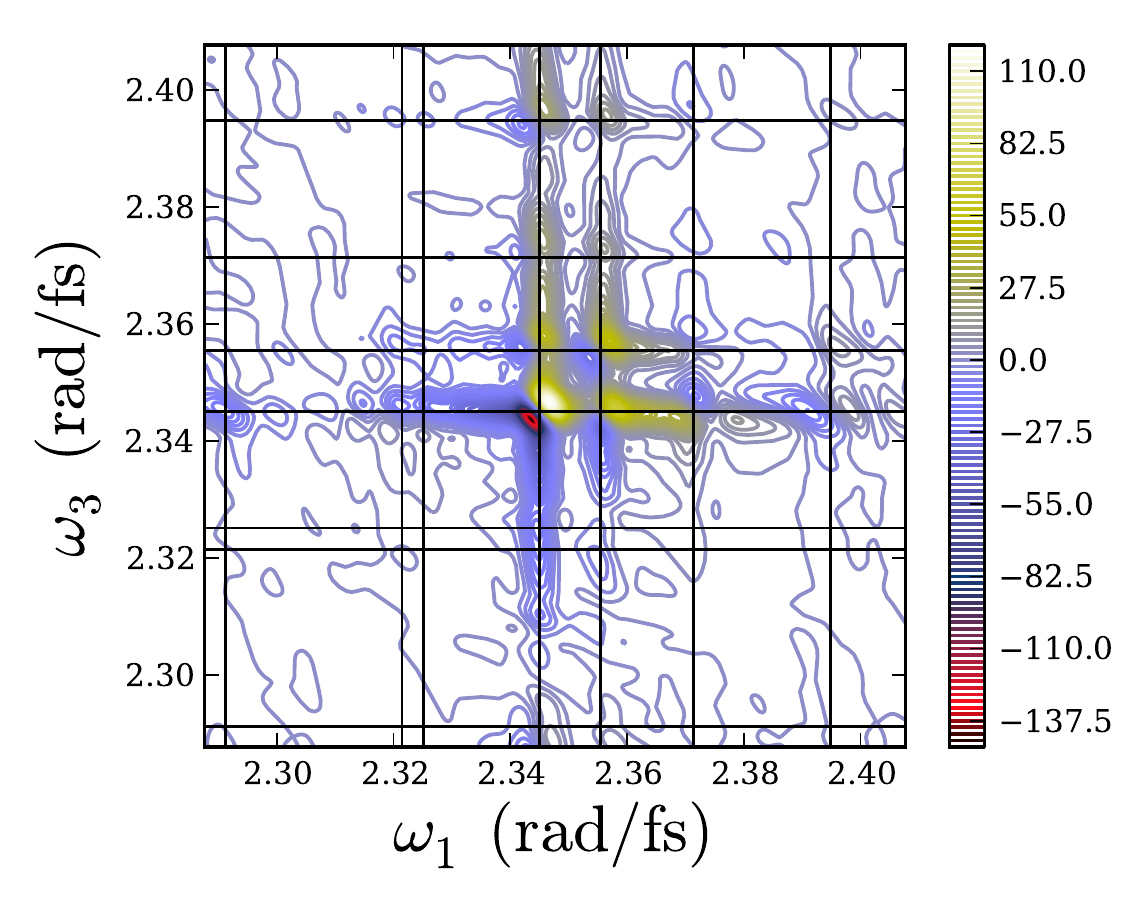} &  \includegraphics[scale=0.6]{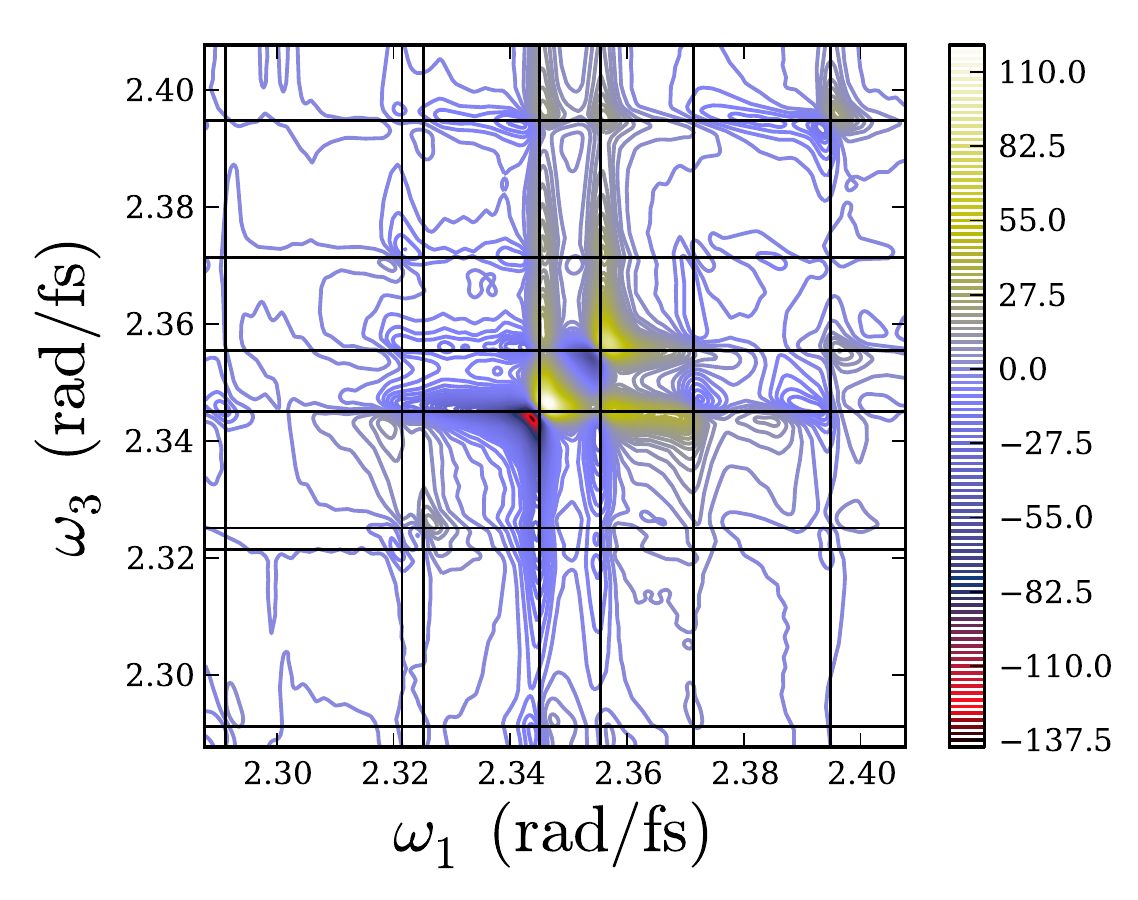}
\end{array}$
\caption{\protect \label{Figent1} The 2-D spectrum of the FMO complex
  with $t_2=0$ (with the parameters described in the text) for a
  random sampling of (topleft) $0.16\%$, (bottomleft) $1\%$,
  (topright) $8\%$ of the total data set $S^{(3)}(t_1,t_2=0,t_3)$ and
  the exact spectrum (bottomright). The black lines indicate the
  eigenfrequencies of the Hamiltonian in the first exciton
  manifold. Even for very small sampling rates the principal features
  of the spectrum are well reproduced and for a sampling of $8\%$ only
  very minor differences remain. Each time variable has $600$ bins
  resulting in a total data set in the form of a $600\times 600$
  matrix. The convergence of the reconstructed matrices $S_k$ to the
  exact spectrum $S_{\rm ex}$ with increasing sampling $k$ can also be
  quantified using the standard Frobenius norm with the magnitude
  $\mathcal{D}_k\equiv \vert\vert S_{\rm ex}-S_k \vert\vert/\vert\vert
  S_{\rm ex}\vert\vert$ which yields $\{\mathcal{D}_k\}_{k=0.16, 1.00,
    8.00}=\{1.81, 0.38, 0.08\}$. For the sake of comparison, the same
  measure averaged over a large sample of entirely random matrices
  $S_{\rm rand}$ yields $\mathscr{D}_{\rm rand}=8.96$. On the other
  hand, the most distant matrix from the exact spectrum gives
  $\mathscr{D}_{\rm worst}=15.72$.}
\end{figure*}

The dynamics of a photo-reactive molecule interacting with a series of
laser beams does not preserve the number of excitations within the
molecule. If we consider the natural assumption that the system is in
its ground state when the first pulse arrives at the sample (i.e.,
$\rho(-\infty)=\vert g\rangle\langle g\vert$), then the only relevant
sectors to compute the third-order response function are the ground
state itself, the one-exciton and the two-excitons manifolds. In FMO,
with seven chromophores, this accounts for $1+7+21=29$ states, which
means that any operator in Hilbert space can be represented as a
matrix of dimensions $29 \times 29$, while a superoperator acting on
the Liouville space can be represented by matrices of dimension
$29^2\times 29^2$. On the other hand, the only operators needed to
compute the response function from eq. \eqref{eq:S3def} are the total
Hamiltonian of the FMO molecule (and the corresponding Liouvillian
$\mathscr{L}_{\rm Ham}$), the dephasing superoperator
$\mathscr{L}_{\rm deph}$ and the total dipolar operator $V$ (with the
corresponding superoperator $\mathscr{V}$). It is indeed a
straightforward procedure to construct these operators in the
manifolds mentioned above \cite{HeinNJP12} and it should be noted that
the only information required to construct them are the physical
matrix elements given by $H_s$ in eq. \eqref{eq:Hsite}, together with
the dipole information contained in eq. \eqref{eq:Dipoles} and the
dephasing rates $\{\gamma_i\}_{i=1\dots 7}$ already written above.

Results of the application of compressive sampling on these 2-D
spectra are presented in Fig.~\ref{Figent1} for population time
$t_2=0$. Already computing only about $0.16\%$ randomly sampled points
(topleft) the position of peaks start to emerge. For higher sampling
ratios of $1\%$ (bottomleft) many features of the spectrum are
reproduced qualitatively even though some minor features are
missing. For $8\%$ (topright) the spectra are reproduced very well
quantitatively and the differences with the exact 2-D spectra
(bottomright) are almost negligible as quantified by the Frobenius norm
difference (see caption of Fig.~\ref{Figent1}). Note that all the spectra have
been normalized to the same maximum value for ease of comparison. The
reconstruction of the $600\times 600$-matrices from the randomly sampled
points was possible within less than $1$ minute on a standard laptop for
$S^{(3)}(t_1, t_2=0, t_3)$ \cite{footnote2}. The full
computation of such a spectrum on the same machine takes around $60$
minutes and scales as $n^2$ for an $n\times n$ matrix.  Assuming a
sampling of $1\%$ of the data set the same calculation could have been
achieved in around $2$ minutes including the SVT reconstruction
implying a $30$ fold improvement in computation time. The relative
efficiency improvement grows almost linearly with $n$. It should also
be noted that 2-D spectra simulations aimed at direct comparison with
experiments typically include rotational and inhomogeneous averages
that would multiply the total computation time by the size of the
sampling distribution used to make each independent averaging. This
does not affect our general conclusion concerning the computational
speedup that can be obtained by means of compressive sampling.

This example shows that even for rich 2-D spectra such as those obtained
for the FMO complex, compressive sampling yields a considerable reduction of the
number of arguments $t_1,t_3$ for the which $S^{(3)}(t_1, t_2=0, t_3)$
needs to be evaluated to obtain the complete 2-D spectrum with a precision
that is sufficient for comparison to experiment. This exemplifies the
power of signal processing concepts to save computational resources.
The extraction of information from complex spectra can also be assisted
by methods from signal processing and in a future publication
we will explore the use of wavelets to this effect.

{\em Conclusions --} We have demonstrated that the principles of
compressed sampling can lead to significant reductions in the
computation of non-linear 2-D spectra.  In such cases only a small
number of data points need to be determined while the rest is
constructed via matrix reconstruction techniques. Our findings are
more general however and can provide considerable computational
savings in a wide variety of computational problems in physics where
data are represented by low-rank matrices.

{\em Acknowledgements --} We acknowledge discussions with T. Baumgratz, F. Caycedo-Soler,
M. Cramer and S.F. Huelga. This work was supported by an Alexander von Humboldt
Professorship, the EU Integrated Project QESSENCE, the BMBF Verbundprojekt QuOReP,
the Ministerio de Ciencia e Innovaci{\'o}n Project No. FIS2009-13483-C02-02
and the Fundaci\'{o}n S\'{e}neca Project No. 11920/PI/09-j.


\begin{thebibliography}{99}
\bibitem{footnote2} This time scale can be improved considerably by making use of the
sparsity of the matrices and algorithmic improvements \cite{MaGC08,BeckerBC11}.
\end{thebibliography}
\end{document}